# Estimation of symbol time offset using cyclic prefix in OFDM system

Ivan Cavlek, Sonja Grgic
University of Zagreb Faculty of Electrical Engineering and Computing
Unska 3, Zagreb 10000, Croatia
ivan.cavlek@fer.hr

*Abstract*—In this paper synchronization techniques using cyclic prefix (CP) are analyzed to remove the influence of symbol time offset (STO) for correct synchronization in Orthogonal Frequency Division Multiplex (OFDM) system. For correct detection of STO using CP two techniques are used. The first one is based on finding the maximum correlation between two blocks and the second one on finding the similarity between two blocks which is maximized when the difference between them is minimized. Two cases are observed. The first one uses only additive white Gaussian noise (AWGN), while the second also uses the channel impulse response (CIR) of Rayleigh channel. When channel effect is added both performances of estimation by correlation and difference deteriorate if sufficient level of signal to noise (SNR) and proper length of CP are not provided.

*Keywords*—Synchronization; OFDM; STO; Cyclic Prefix

## I. INTRODUCTION

In modern communication systems, the main goal of research and development is to achieve the highest possible transmission rates in channels with limited frequency bandwidth and to improve the quality of service and the performance of the system [1, 2]. In wireless communication two basic techniques are used: single-carrier transmission and multicarrier transmission. In single-carrier systems one transmission frequency is used to transmit data (serial data transmission). The transmitted symbols had a duration of $T$ seconds; therefore, the data rate was $R = 1/T$. They were sent through a channel with characteristic $h(t)$. Before being sent they had to be pulse-shaped (due to spectrum limitations) by transmitting filter $g_T(t)$ in the transmitter, and at the receiver processed with the receiver filter ($g_R(t)$) and the equalizer ($h^{-1}(t)$). Also, AWGN noise ($z(t)$) would be added in the channel [3]. The pictorial representation is shown in Figure 1.

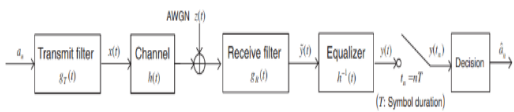

Figure 1. Single-carrier communication system model

To increase the data rate, we could shorten the symbol time or use higher order modulation procedures. The number of discrete states is limited by noise, and short symbol time causes the larger influence of inter-symbol interference (ISI), and the complexity of equalizers rises. As a solution to these problems, a new system was designed. Instead of single-carrier transmission, multiple carriers are used (parallel transmission). In multicarrier systems frequency channel is divided into subcarriers and high-rate data stream is divided into low-rate data substreams that are parallelly transmitted over subcarriers. This method is known as Orthogonal Frequency Division Multiplex (OFDM). A wideband signal is divided into multiple narrowband signals (each having its own subcarrier frequency). As these narrowband channels are now frequency – nonselective the complexity of equalizers is reduced. A very important property is maintaining the orthogonality between subcarriers to avoid inter-carrier interference (ICI). To produce orthogonal subcarriers Discrete Fourier Transform (DFT) is used [3]. The OFDM transmission scheme is displayed in Figure 2.

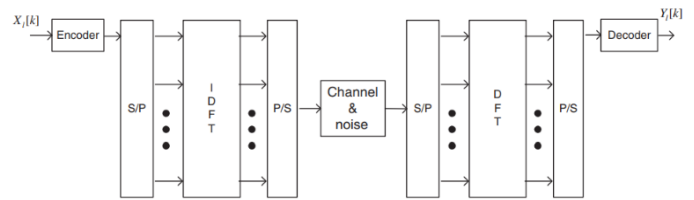

Figure 2. OFDM transmission scheme using IDFT/DFT

To maintain orthogonality, subcarriers are spaced at $1/T$. Spectrum of OFDM signal can be considered as the sum of frequency shifted sync functions in frequency domain. To counteract the effect of ISI a guard interval in time domain, which is usually called cyclic prefix (CP), is inserted in the OFDM scheme. OFDM most often use PSK or QAM as modulation procedures [5]. The potential problem in OFDM can be caused by effects of symbol time offset (STO) and carrier frequency offset (CFO). To properly detect the starting point of OFDM symbol, symbol-timing synchronization must be performed. STO synchronization techniques can be performed in time or frequency domain. Realization through time domain is called coarse symbol timing synchronization, and through frequency domain fine symbol timing synchronization. One of the possible algorithms to solve STO problem is MLE (Maximum Likelihood Estimation) [6]. MLE works very well in AWGN channel, but its performances are degraded when effects of CFO are present and if channel conditions deteriorate [7]. Another possible algorithm that is used is MAP (maximum a-posteriori). The disadvantage of MAP algorithm is that it gives good results for CFO estimation, but not for STO [4]. In this paper we will analyze STO estimation in time domain using CP. Two methods will be applied: correlation-based method (CBM) and difference-based method (DBM).

## II. POSSIBLE TRANSMISSION PROBLEMS

There are four different cases of timing offset, comparing the estimated starting point and exact timing instance. The first case is when the estimated starting point coincides with the exact timing instance. In this case orthogonality is preserved and therefore OFDM symbol can be recovered. In the second case the estimated starting point is before the exact timing instance, but after the end of maximum delay of the previous OFDM symbol (ISI has no influence here, but rotation around the origin in the constellation diagram happens due to phase offset). The following case is when the estimated starting point overlaps with the channel response of the previous OFDM symbol, causing ISI and ICI. The final case is when the estimated starting point is after the exact timing instance (ISI and ICI also occur here) [3, 8]. A graphic representation of the previously described cases is shown in Figure 3.

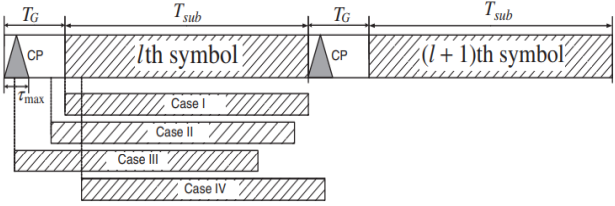

Figure 3. Four possible cases for STO [3]

Before the baseband signal is sent, it is modulated on carrier frequency and at the receiver converted down by the same carrier frequency. One problem that occurs is phase noise caused by instability of carrier signal generators at transmitter ($f_c$) and receiver ($f'_c$). The other is a consequence of Doppler frequency shift, and it is called CFO. Doppler frequency ($f_d$) is given by the following expression $f_d=(v \cdot f_c)/c$ [3] where $v$ is speed of receiver and $c$ is speed of light.

## III. MATHEMATICAL DESCRIPTION OF THE PROBLEM

As previously mentioned, STO may cause phase distortion and ISI. Phase distortion can be compensated by equalizer, but ISI can't be corrected. Although estimation using training symbols gives better results, we will analyze STO estimation using CP. There are two possibilities for correct detection of STO using CP. The correlation-based method (CBM) is set up on finding the maximum correlation between two blocks (purpose of blocks will be described later in text), and difference-based method (DBM) is set up on finding the similarity between two blocks, which is maximized when the difference between them is minimized. Both STO estimation by correlation and difference will be used. Since the cyclic prefix is a copy of a part of the symbol (most often the end), this implies that we could look for similarities for estimation. We will use previously mentioned blocks, which are the same length as CP (has $N$ samples over $T$ seconds). The blocks can move, to find the similarity between the samples. Maximum similarity is obtained when the CP of OFDM symbol is found in the first moving block. So, the first method finds STO by searching the point where the squared difference (we use squared difference to deal with CFO) between two blocks is minimized [3]. The mathematical formulation is given by:

$$\hat{\delta} = arg\,\min_{\delta}\{\sum_{i=\delta}^{N_G-1+\delta}(|\,y_l[n+i] - y_l^*[n+N+i]\,|)^2\} \quad (1)$$

The second approach is to look for correlation between the two blocks. The mathematical formulation is given by:

$$\hat{\delta} = arg\,\max_{\delta}\{\sum_{i=\delta}^{N_G-1+\delta} y_l[n+i]y_l^*[n+N+i]\} \quad (2)$$

In the above equations $\hat{\delta}$ represents the estimated time delay, $\delta$ actual time delay, $N_G$ number of samples in CP, $y_l[n+i]$ received signal at time $n+i$ for the $l$-th receive antenna, $y_l^*[n+N+i]$ complex conjugate of received signal at time $n+N+i$ for the $l$-th receive antenna and $N$ is the number of symbols transmitted in the signal. The two equations are general equations for multiple-input multiple-output (MIMO) system. So, if there are multiple antennas at the receiver, each antenna will have its own index $l$. Both performances are degraded when effects of CFO are added.

## IV. RESULTS AND DISCUSSION

The code for the analysis of STO estimation using CP is implemented by script written in Matlab. In the first case generated OFDM symbols will be sent through the channel with AWGN, and in the second case will include channel effects (using channel impulse response). In this paper we will ignore the effects of CFO. Results will be displayed depending on the level of the signal-to-noise (SNR) ratio and the length of CP. We will examine the behavior of the system when SNR is 10 dB and 2 dB, while the length of CP will amount to 32 and 16 samples. SNR value of 10 dB provides good transmission conditions while SNR value of 2 dB makes transmission system sensitive to deteriorations. A similar situation is valid for CP. The length of CP of 32 samples provides good synchronization conditions while the length of CP of 16 samples makes the system sensitive to synchronization errors.

Graphical representations of the results for the first case are shown in Figures 4-7. The red and green plots represent the performance of correlation-based method (green) and difference-based method (red). The x-axis shows sample number, and the y-axis represents the magnitude of the correlation or difference value. Blue line represents the maximum correlation value, red line minimum difference value and black dot true STO value. In Figures 4 and 5 length of CP is constant (32 samples) and SNR is degraded from 10 dB to 2 dB. When the value of SNR is 10 dB algorithm works fine and can detect where STO happened. When the value is reduced to 2 dB STO can no longer give reliable information where offset happened. Figures 6 and 7 show that if the length of CP is decreased (16 samples), we get even sooner for the same values of SNR (10 dB and 2 dB) unsatisfactory results. Table I shows estimated values of STO for CBM and DBM in AWGN channel.

In the second case channel effects will be included. We will use the channel impulse response (CIR) whose values are obtained from Gaussian random variable with zero mean and unit variance (we are modeling Rayleigh fading channel). The

CIR will consist of 10 coefficients which are different every time we run code (because we are using rand function). The values of the coefficients are respectively [-0.2338 + 0.1770i, 0.1573 - 0.0179i, 0.1352 + 0.1641i, -0.1318 - 0.2919i, -0.1715 + 0.3104i, 0.5049 - 0.1209i, 0.2021 - 0.6263i, 0.0621 + 0.1324i, 0.1568 - 0.0362i, 0.0113 - 0.0004i]. In theory, such channels should degrade the performances of our estimation algorithms. Magnitude and phase response of such channel is shown in Figure 8. Magnitude response shows the gain or attenuation of the filter at each frequency, while the phase response shows the amount of phase shift introduced by the filter at each frequency.

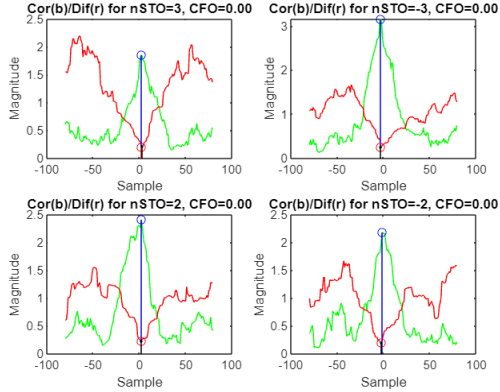

Figure 4. Estimated values of STO using CBM (green) and DBM (red) (SNR = 10 dB, CP = 32)

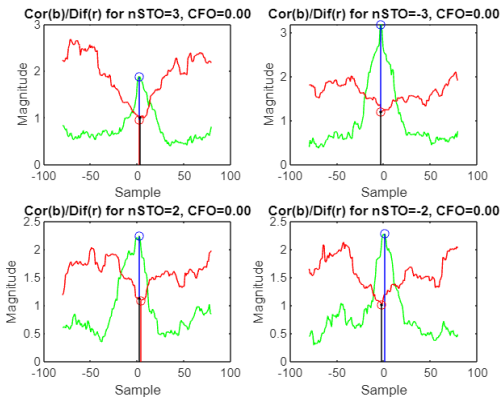

Figure 5. Estimated values of STO using CBM (green) and DBM (red) (SNR = 2 dB, CP = 32)

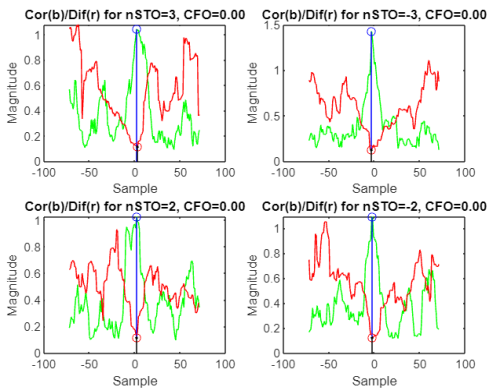

Figure 6. Estimated values of STO using CBM (green) and DBM (red) (SNR = 10 dB, CP = 16)

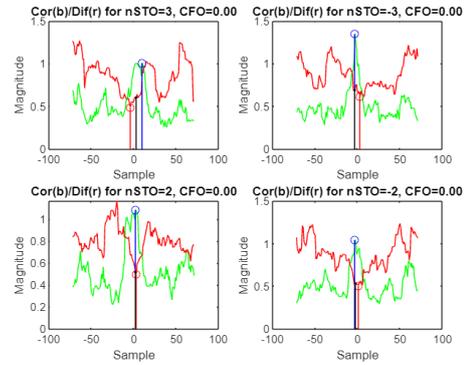

Figure 7. Estimated values of STO using CBM (green) and DBM (red) (SNR = 2 dB, CP = 16)

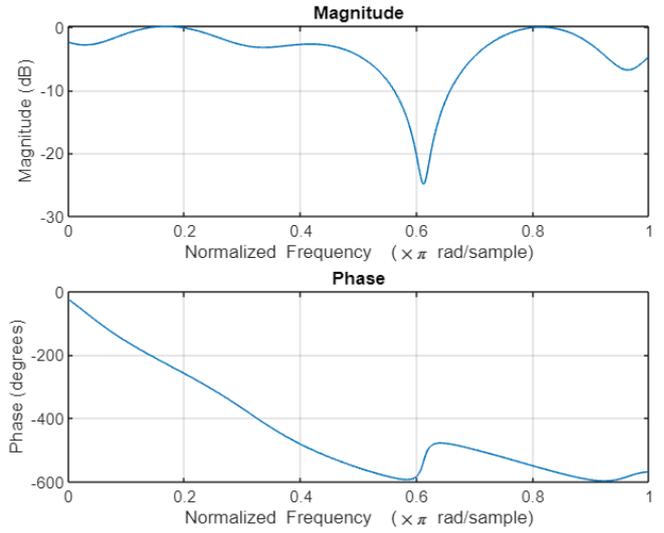

Figure 8. Magnitude and phase response of complex-valued filter

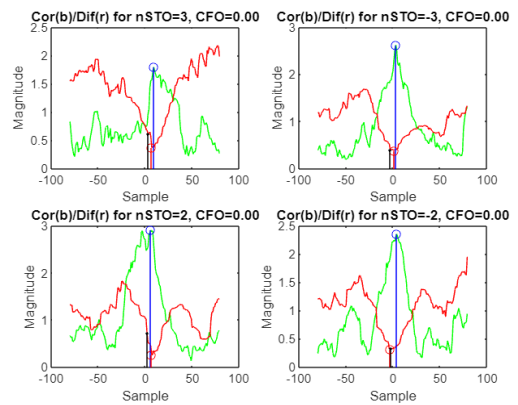

Figure 9. Estimated values of STO using CBM (green) and DBM (red) (SNR = 10 dB, CP = 32, CIR included)

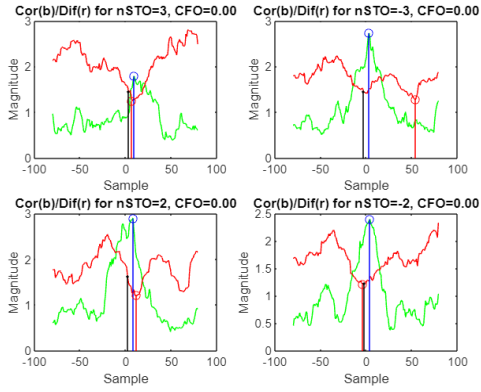

Figure 10. Estimated values of STO using CBM (green) and DBM (red) (SNR = 2 dB, CP = 32, CIR included)

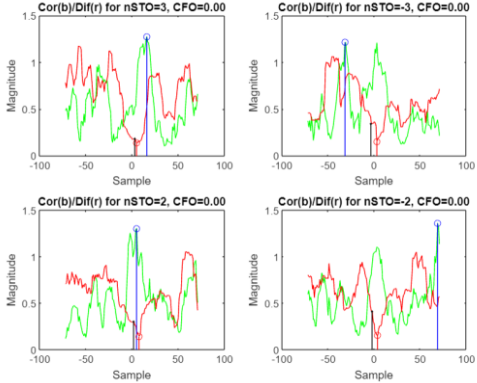

Figure 11. Estimated values of STO using CBM (green) and DBM (red) (SNR = 10 dB, CP = 16, CIR included)

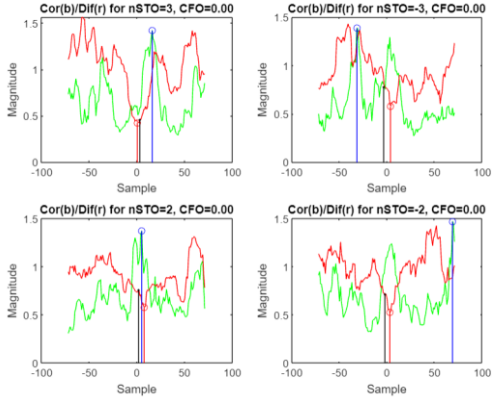

Figure 12. Estimated values of STO using CBM (green) and DBM (red) (SNR = 2 dB, CP = 16, CIR included)

Figures 9-11 represent what happens when channel effect is added. Table II shows estimated values of STO for CBM and DBM in Rayleigh fading channel. Comparing Figure 4 and Figure 9 the efficiency of the algorithm has dropped. In Figure 4, the estimated values correspond to the actual values of STO (3, -3, 2, -2) for DBM but for CBM are slightly different (2, -3, 2, -1). Figure 9 shows how all three lines (black, blue, and red) diverge. Their values are respectively (3, 9, 7), (-3, 3, 1), (2, 6, 7) and (-2, 4, -3). In such conditions the error increases which means that algorithms can no longer properly estimate the positions of STO. In the worst cases (Figure 11 and 12), huge deviations can be seen for correlation method (values -31 and 69, instead of -3 and -2).

TABLE I. ESTIMATED VALUES OF STO USING CBM AND DBM IN AWGN CHANNEL

| SNR | True STO | CP=32 | | CP=16 | |
| --- | --- | --- | --- | --- | --- |
| | | CBM STO | DBM STO | CBM STO | DBM STO |
| 10 dB | 3 | 2 | 3 | 2 | 3 |
| | -3 | -3 | -3 | -3 | -3 |
| | 2 | 2 | 2 | 2 | 2 |
| | -2 | -1 | -2 | -2 | -2 |
| 2 dB | 3 | 2 | 2 | 10 | -4 |
| | -3 | -3 | -3 | -3 | 3 |
| | 2 | 2 | 4 | 2 | 3 |
| | -2 | 1 | -2 | -3 | 1 |

TABLE II. ESTIMATED VALUES OF STO USING CBM AND DBM IN RAYLEIGH FADING CHANNEL

| SNR | True STO | CP=32 | | CP=16 | |
| --- | --- | --- | --- | --- | --- |
| | | CBM STO | DBM STO | CBM STO | DBM STO |
| 10 dB | 3 | 9 | 7 | 16 | 5 |
| | -3 | 3 | 1 | -31 | 3 |
| | 2 | 6 | 7 | 5 | 8 |
| | -2 | 4 | -3 | 69 | 4 |
| 2 dB | 3 | 9 | 7 | 16 | 1 |
| | -3 | 3 | 54 | -31 | 4 |
| | 2 | 8 | 12 | 5 | 8 |
| | -2 | 4 | -4 | 69 | 3 |

V. CONCLUSION

Estimation of STO has been presented for OFDM system using CP. Two methods were being analyzed. The effects of the methods were observed depending on the length of CP, level of the SNR and the addition of CIR. How the SNR level decreases, it is increasingly difficult to estimate where STO happened. Both estimation by correlation and difference are ineffective if a sufficient level of SNR and appropriate length of CP is not provided. The only way to eliminate this problem is to always ensure a sufficient level of SNR and appropriate length of CP or use some more complex techniques.